\documentclass[pre,twocolumn,aps,superscriptaddress,showpacs,floatfix]{revtex4}
\usepackage{graphicx}
\usepackage{dcolumn}
\usepackage{bm}

\begin{document}
\title{Geometric stochastic resonance in a double cavity}

\author{Pulak K. Ghosh}
\affiliation{Advanced Science Institute, RIKEN, Wako-shi, Saitama, 351-0198, Japan}
\author{Russell  Glavey}
\affiliation{Department of Physics, Loughborough University,
Loughborough LE11 3TU, United Kingdom}
\author{Fabio Marchesoni}
\affiliation{Dipartimento di Fisica, Universit\`a di Camerino,
I-62032 Camerino, Italy}
\affiliation{Advanced Science Institute, RIKEN, Wako-shi, Saitama, 351-0198, Japan}
\author{Sergey E. Savel'ev}
\affiliation{Department of Physics, Loughborough University,
Loughborough LE11 3TU, United Kingdom}
\affiliation{Advanced Science Institute, RIKEN, Wako-shi, Saitama, 351-0198, Japan}
\author{Franco Nori}
\affiliation{Advanced Science Institute, RIKEN, Wako-shi, Saitama, 351-0198, Japan}
\affiliation{Department of Physics, University of Michigan, Ann
Arbor, MI 48109-1040, USA}
\date{\today}
\begin{abstract}
Geometric stochastic resonance of particles diffusing across  a
porous membrane subject to oscillating forces is characterized as a
synchronization process. Noninteracting particle currents through a
symmetric membrane pore are driven either perpendicular or parallel
to the membrane, whereas harmonic-mixing spectral current components
are generated by the combined action of perpendicular and parallel
drives. In view of potential applications to the transport of
colloids and biological molecules through narrow pores, we also
consider the role of particle repulsion as a controlling factor.
\end{abstract}

 \maketitle

\section{Introduction} \label{intro}
\begin{figure}[tp]
\centering
\includegraphics[width=8.1cm]{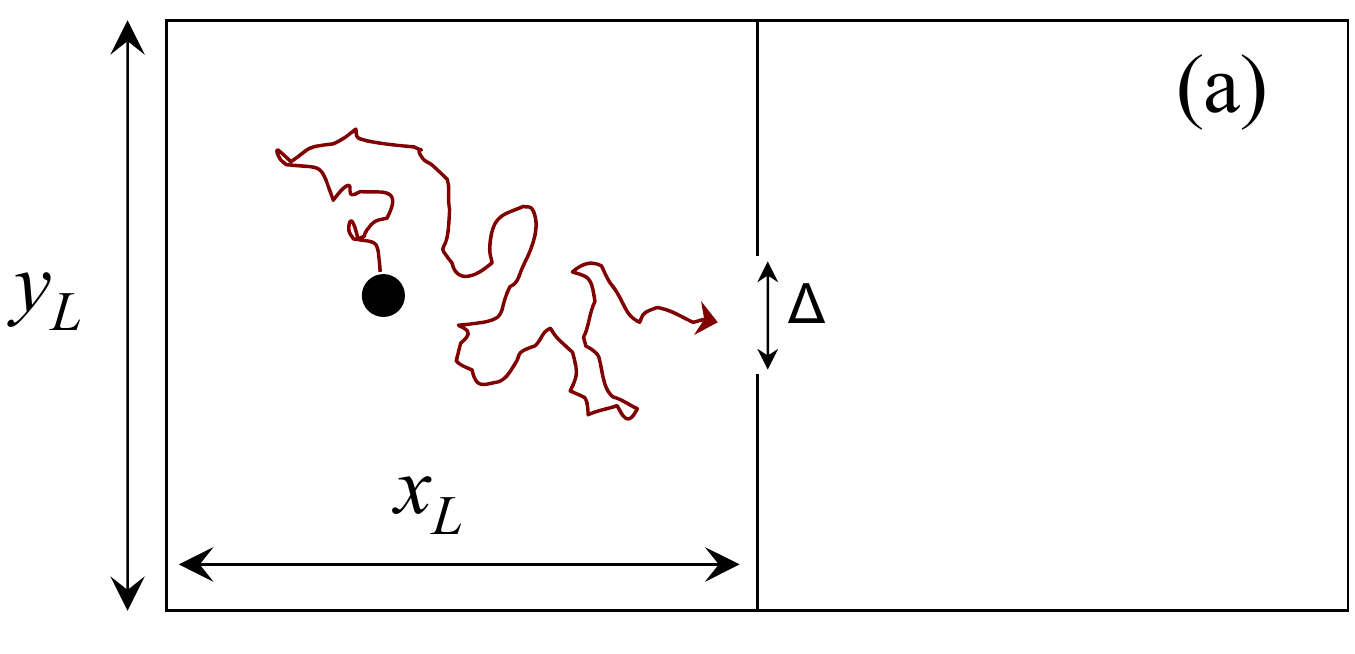}
\includegraphics*[width=8.2cm]{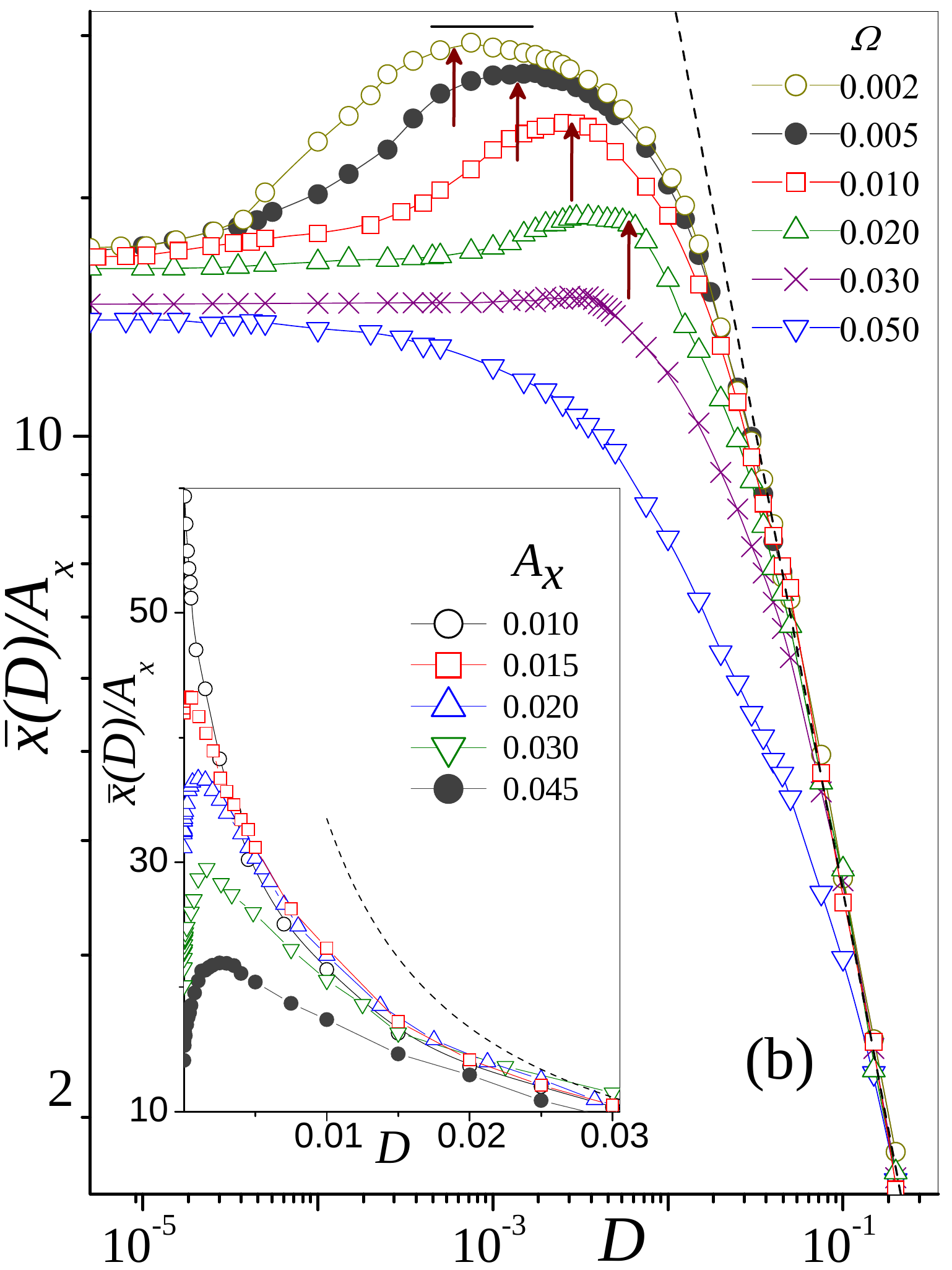}
\caption{(Color online) (a) Brownian particle confined to a 2D box
divided in two compartments by a partition with an opening of width
$\Delta$ at the center. (b) Geometric stochastic resonance.
$\overline x(D)$ versus $D$ for different values of $\Omega$ at
$A_0=0.045$ (main panel) and $A_x$ at $\Omega=0.01$ (inset). Other
parameters are: $x_L=y_L=1$ and $\Delta =0.1$. In the inset we also
report our predictions for the SR peak position $D_{\rm max}$
(vertical arrows) and height $\overline x(D_{\rm max})$ (top line).
In both plots the dashed line represents the decay law $\overline
x(D\to \infty)$ (see Sec. \ref{gSR}). \label{F1}}
\end{figure}

Historically, research on Stochastic Resonance (SR)~\cite{RMP}
focused mostly on systems with purely energetic potentials, either
continuous or discrete. However, as pointed out in
Ref.~\cite{Reguera}, in soft condensed matter and in a variety of
biological systems \cite{natural,BM}, particles are often confined
to constrained geometries, such as interstices, pores, or channels,
whose size and shape can affect the SR mechanism \cite{Burada1}.
Indeed, smooth confining geometries can be modeled as entropic
(i.e., noise or temperature dependent) potentials \cite{Zwanzig},
capable of influencing the response of the system to an external
driving force (see, for a review, Ref. \cite{chemphyschem}).

A more interesting scenario emerges in the case of  sharp confining
geometries. In a recent paper \cite{PRLgSR} we showed that a
Brownian particle confined to two distinct cavities divided by a
porous medium, say a membrane, undergos SR when driven by an ac
force perpendicular to the membrane. This means that, at variance
with ordinary SR \cite{RMP}, optimal synchronization between drive
and particle oscillations for an appropriate noise level, occurs
even in the \emph{absence} of a bistable effective potential, of
either energetic or entropic nature \cite{Burada1}. However, such a
manifestation of SR in higher dimensions requires adopting extremely
sharp geometrical constrictions to separate the two cavities. The
magnitude and conditions of such a {\it geometric} SR effect are
sensitive to the geometry of the cavities and the structure of the
pores.

This instance geometric SR should not be mistaken for the socalled
{\it entropic} SR  introduced in Ref.~\cite{Burada1}, where the
Brownian particle can switch cavity only by overcoming the entropic
barrier determined by the geometric constriction associated with a
smooth pore. In the absence of an energetic barrier, the entropic
barrier alone determines the magnitude of the SR effect that occurs
when a periodic force drives the particle across the pore, although
the evidence reported by Burada {\it et al.}~\cite{Burada1} hints at
an interplay of entropic and energetic barriers, rather than to a
mere entropic effect.

Geometric SR is a peculiar manifestation of driven Brownian motion
in septate channels \cite{bere1,bere2,lboro1,PHfest,shape}. In these
channels compartments are separated by zero-thickness partition
walls and the pores are modeled by structureless holes pierced at
the center of the partition walls. A septate channel cannot be
analyzed in terms of the reductionist approach of Ref.
\cite{Zwanzig}, i.e., as entropic channels
\cite{Reguera,chemphyschem}, because the geometry of its pores turns
out to be too sharp for a one dimensional (1D) kinetic equation
approximation to hold \cite{Jacobs}. However, sharp pore geometries
are known to enhance most of the noise controlled transport
mechanisms proposed in recent years \cite{BM}, thus making this
class of channels particularly suitable for experimental
demonstrations.

In this paper we restrict our analysis to two dimensional (2D)
geometries.  Our choice is partly motivated by the growing interest
in vortex superconducting devices, a class of artificial devices
with numerous technological applications, including flux qubits,
SQUIDs and superconducting rf filters \cite{vortexzhou,vortex}. Superconducting
samples with two vortex boxes connected by a thin pore of almost any
geometry can be easily fabricated. Vortices are trapped inside the
boxes with binding energy of the order of $\Phi_0^2L_t/\lambda^2$,
where $\Phi_0$ is the magnetic flux quantum, $\lambda$ is the London
penetration depth, and $L_t$ is the depth of the two vortex traps.
Magnetic vortices repel one another through a logarithmic pair
potential, while their density, $n=H/\Phi_0$, is controlled by the
intensity $H$ of the applied magnetic field. In the dilute limit,
$H\lesssim \Phi_0/\lambda^2$, the vortex-vortex interactions become
negligible, so that the transport properties of a single trapped
vortex are not overshadowed by many-body effects. ac drives and
noise sources can be easily implemented as Lorentz forces generated
by independent electric currents injected into the sample parallel
and perpendicular to the pore axis. Detection of SR under such
experimental conditions is thus regulated by the applied current
sources only; in particular, the noise parameter can be varied
independently of the constant operating sample temperature.

This paper is organized as follows. In Sec. \ref{gSR}  we summarize
the spectral properties of geometric SR, as first reported in Ref.
\cite{PRLgSR}. In Sec. \ref{sync} we analyze geometric SR as a noise
controlled synchronization mechanism. The distribution of the
crossing times of a single particle through a pore is investigated
in the case of both longitudinal and transverse ac drives. In Sec.
\ref{HM} we study the particle current through a pore under the
combined action of longitudinal and transverse drives with
incommensurate frequencies. Harmonic mixing is detected as a
consequence of confinement, even in the absence of nonlinear
particle interactions. In Sec. \ref{interaction} we explore the role
of particle repulsion in controlling geometric SR, by varying the
particle density and the pair interaction intensity. Finally, in
Sec. \ref{conclusion} we add some concluding remarks.

\section{Geometric stochastic resonance} \label{gSR}

Let us consider an overdamoed Brownian particle freely diffusing in
a 2D suspension fluid contained in two symmetric $x_L \times y_L$
compartments, with reflecting walls \cite{gardiner}, connected by a
narrow pore of width $\Delta$, as illustrated in Fig. \ref{F1}(a).
The overdamped dynamics of the particle is modeled by the Langevin
equation,
\begin{equation}\label{le}
\frac{d{\vec r}}{dt}=-{\vec A}(t)\;+ \sqrt{D}~{\vec \xi}(t),
\end{equation}
where ${\vec A} = (A_x,A_y)$ are the $x, y$ components of the
driving force and ${\vec \xi}(t)=(\xi_x(t),\xi_y(t))$ are zero-mean,
white Gaussian noises with autocorrelation functions $\langle
\xi_i(t)\xi_j(t')\rangle = 2\delta_{ij}\delta(t-t')$, with
$i,j=x,y$. Equation (\ref{le}) has been numerically integrated by a
Milstein algorithm \cite {milstein}. Stochastic averages were
obtained as ensemble averages over 10$^6$ trajectories with random
initial conditions; transient effects were estimated and subtracted.

In the presence of a longitudinal ac drive oriented along  the $x$
axis, $A_x(t)=A_x\cos (\Omega t)$, the Brownian trajectories embed a
persistent harmonic component, $\overline x(D) \cos[\Omega t
-\phi(D)]$, whose amplitude, $\overline x$ is plotted versus $D$ in
Fig.~\ref{F1}(b). For details about the curves $\phi(D)$, not
relevant to the discussion below, the reader is referred to the
original paper \cite{PRLgSR}.

\begin{figure}[bp]
\centering
\includegraphics*[width=8.2truecm]{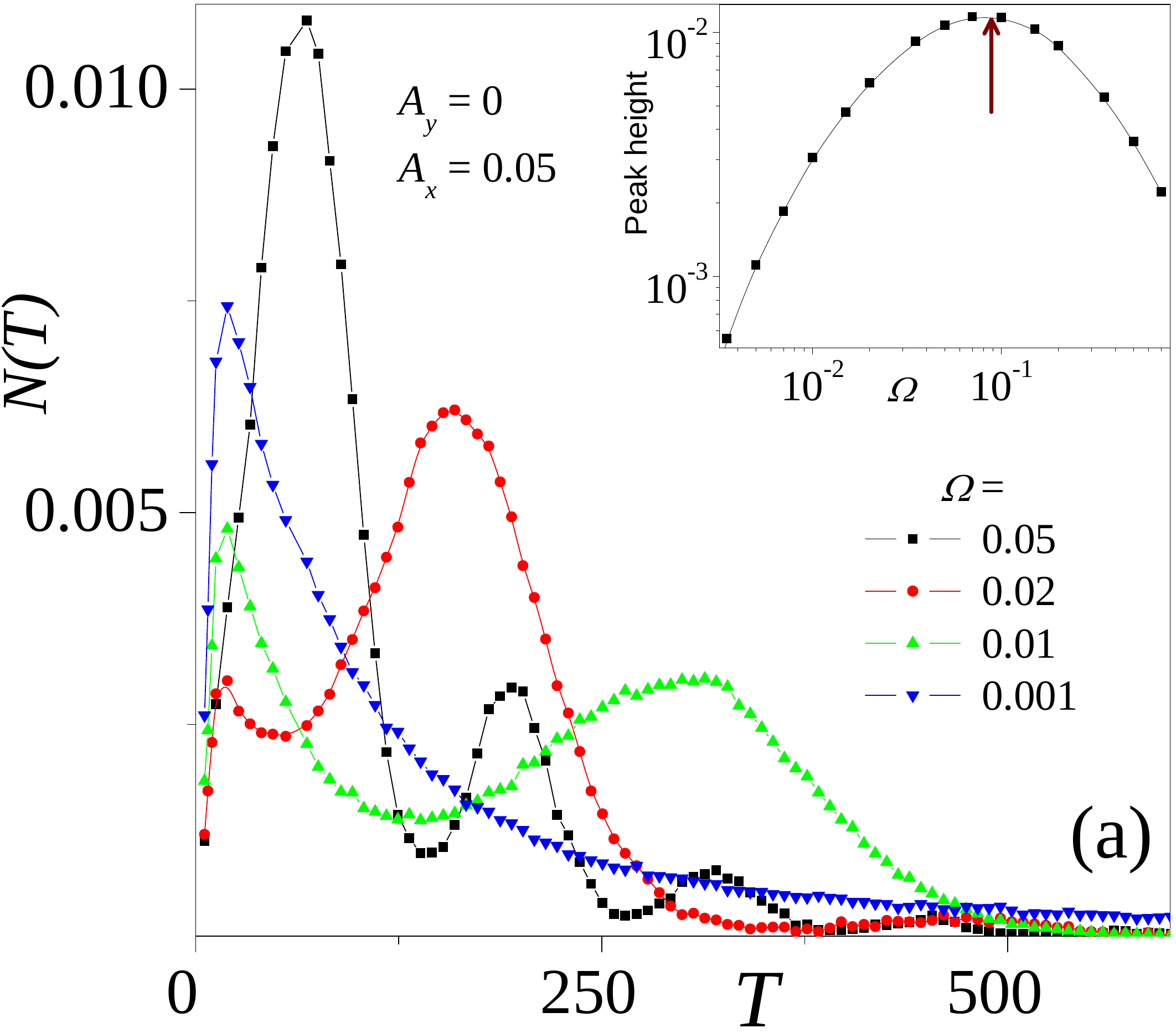}
\includegraphics*[width=8.2truecm]{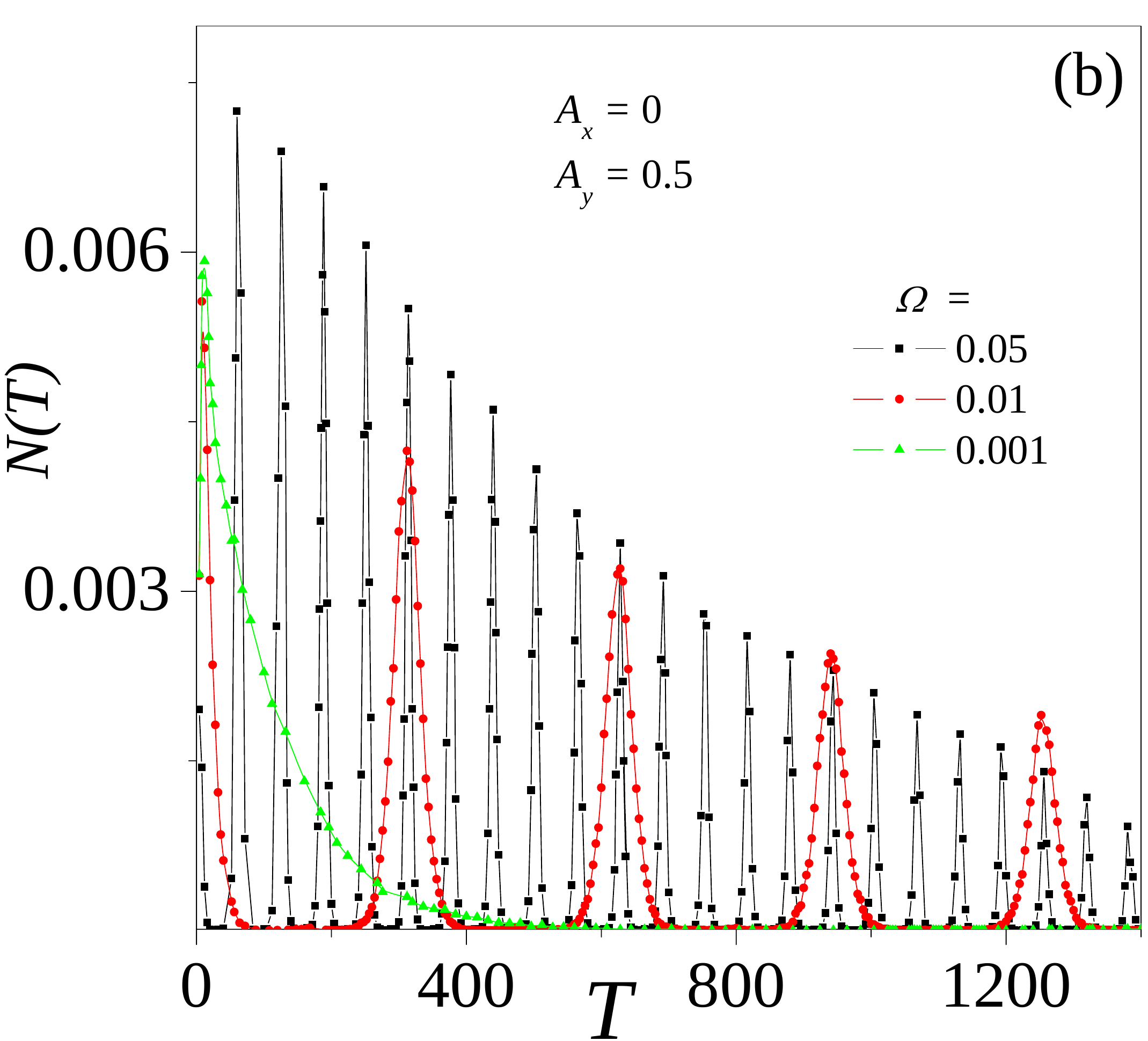}
\caption{(Color online) Synchronization mechanism. Distribution
density of the residence times for different periods
$T_\Omega=2\pi/\Omega$ (in the legends) of the (a) longitudinal
drive with $A_x=0.05$ and (b) transverse drive with $A_y=0.5$. The
cavity dimensions are as in Fig. \ref{F1} and $D=0.015$. Inset in
(a): height of the first peak, $n=1$, versus $\Omega$. Vertical
arrow denotes the resonant $\Omega$ value predicted in Eq.
(\ref{bf}) with $\tau_\Delta$ given in \cite{PHfest}. \label{F2}}
\end{figure}

The occurrence of geometric SR is clearly  documented in Ref.
\cite{PRLgSR}. Its main distinctive properties that can be
summarized as follows:

(i) $\overline x(D)$ peaks for an appropriate noise intensity, $D_{\rm max}$, with upper bound,
\begin{equation}
\label{xmax}
\bar x(D_{\rm max}) \leq \frac{4}{\pi}\;x_L.
\end{equation}
A satisfactory estimate of $D_{\rm max}$ was obtained by matching
the half-drive period with mean first-exit time from one compartment
at $A_x=0$, $\tau_{\Delta}(D)$, that is,
\begin{equation}
\label{Dmax}
\tau_{\Delta}(D_{\rm max}) = \frac{T_\Omega}{2}\equiv\frac{\pi}{\Omega}.
\end{equation}
Note that $\tau_{\Delta}(D)$  is inversely proportional to $D$ and
strongly depends on the pore width, $\Delta$, as computed in Ref.
\cite{PHfest}. Hence, $D_{\rm max}$ is proportional to $\Omega$;

(ii) SR is restricted to $A_0>A_c$ or $\Omega < \Omega_c$, as a consequence of the geometric condition,
\begin{equation}
\label{A_c}
\frac{A_x}{\Omega}\geq \frac{4}{\pi}\;x_L.
\end{equation}
This is an important difference with respect  to ordinary SR
\cite{RMP}, where there exist no such onset thresholds in the drive
parameters space;

(iii) $\overline x(D)$ obeys the approximate SR curve,
\begin{equation}
\label{xvsD}
\bar x(D)=\frac{\bar x_{0}(D)}{\sqrt{1+ [\Omega\tau_\Delta(D)]^2}},
\end{equation}
where,
\begin{equation}
\label{new}
\bar x_{0}(D)=\frac{{A_x x_L}/{D}}{\tanh\left ({A_x x_L}/{D}\right )}-1,
\end{equation}
is the amplitude of the $\langle x(t) \rangle$  oscillations in the
adiabatic limit $\Omega \to 0$. As a consequence, for large noise,
$\bar x(D) \propto 1/D$.

The properties listed above follow from simple geometrical
considerations.  Equation (\ref{Dmax}) is the
\emph{geometric counterpart of the standard SR condition}~\cite{RMP}, where, in the absence of an
 energetic barrier, the Arrhenius time is replaced by an appropriate diffusion time across the pore.
  As is apparent in Eq.~(\ref{new}), for weak noise and low drive frequencies, $A_x(t)$ presses the particle against the walls of the container opposite to the dividing wall and, as a result, the average particle displacement, $\langle x(t) \rangle$, approaches a square waveform with amplitude $x_L$. 
The amplitude of the Fourier component of $\langle x (t) \rangle$ with angular frequency $\Omega$ is then, $4x_L/\pi$.
From this remark it immediately follow the inequalities in
Eqs.~(\ref{xmax}) and (\ref{A_c}), $A_x/\Omega$ being the driven
oscillation amplitude of an unconstrained Brownian particle. The
upper bound of Eq.~(\ref{xmax}) hold for vanishingly low $\Omega$,
see Fig.~\ref{F1}(b). For stronger noise but still low drive
frequencies, $\bar x_0(D)$ tends to $A_x \tau_x$, where
$\tau_x=x^2_L/3D$ is the longitudinal diffusion time across a cavity
compartment. The pore effectively suppresses the particle
oscillations, with damping constant $\tau_\Delta(D)^{-1}$, only at
relatively high drive frequencies, as apparent from
Eq.~(\ref{xvsD}). The consistency of these analytical results with
the simulations is quite satisfactory, as shown in~\cite{PRLgSR}.

\begin{figure}[bp]
\centering
\includegraphics*[width=8.0truecm]{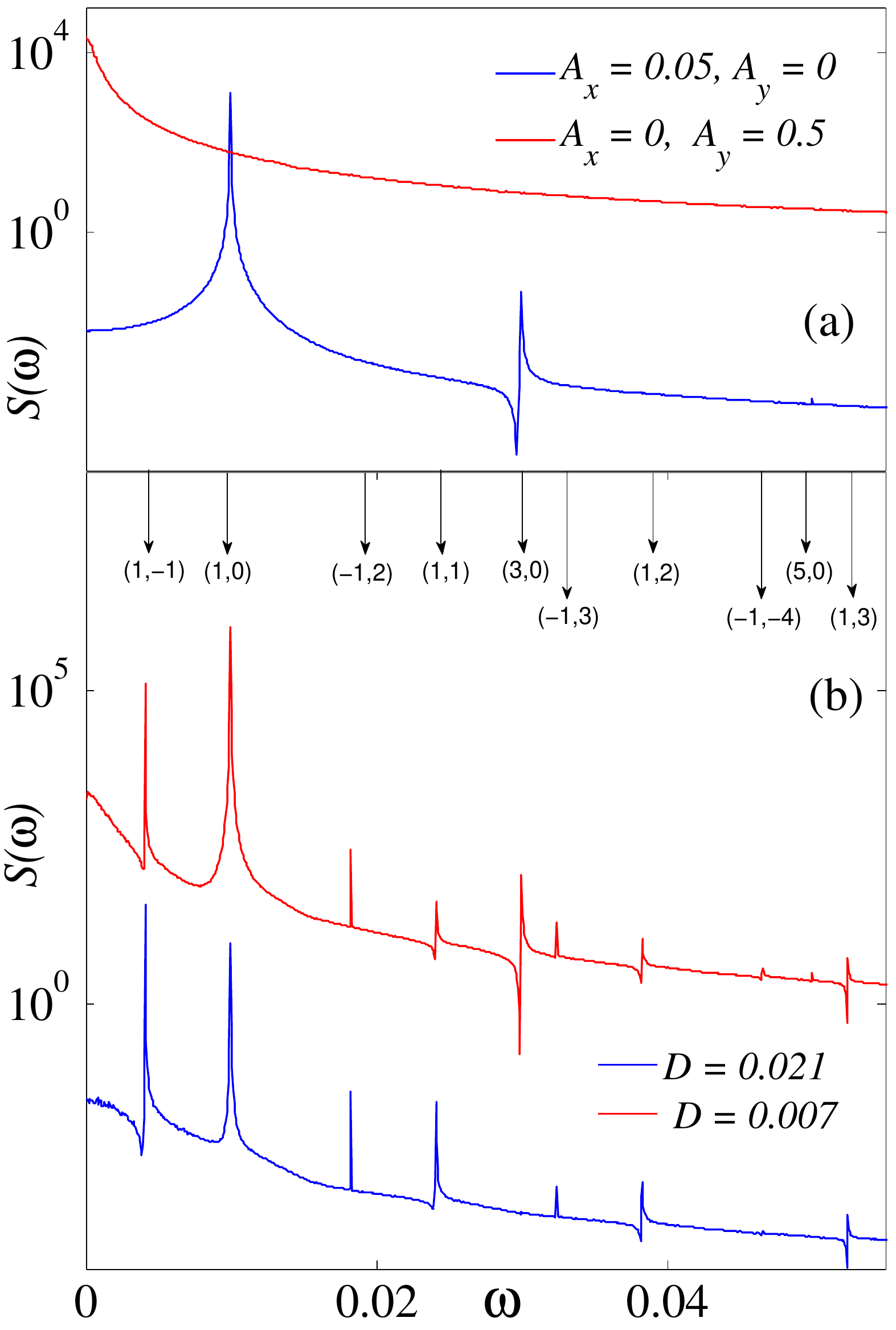}
\caption{(Color online) Harmonic mixing in a  cavity. Spectral
density of the coordinate $x(t)$ for two different noise
intensities, $D$, and (a) either a longitudinal or a transverse
drive (see legend); (b) a combination of longitudinal and transverse
drives. The drive parameters are $A_x=0.05$, $A_y=0.5$,
$\Omega_x=0.01$, and $\Omega_y=\Omega_x/\sqrt{2}$. The harmonic
mixing resonances are indicated by down-pointing vertical arrows
with relevant indices $(m,n)$. \label{F3}}
\end{figure}

\section{Synchronization mechanisms} \label{sync}

As summarized in Sec.~\ref{gSR}, the evidence for geometric SR
reported in Ref. \cite{PRLgSR} focuses on the $D$ dependence of the
harmonic component of $\langle x(t)\rangle$ with driving frequency
$\Omega$. This spectral characterization of geometric SR will be
further discussed in the following section.

An alternative and perhaps deeper insight  into the underlying
resonance mechanism can be gained by considering the so-called
synchronization characterization of SR \cite{gammaitoni}. In this
approach one looks at the residence times, $T$, of the Brownian
particle in either cavity -- which one is irrelevant, due to the
mirror symmetry of the process with respect to the cavity divide.
Their distribution densities, $N(T)$, exhibit a prominent peak
structure and, more remarkably, a resonating $D$ and $T$ dependence.

\subsection{Longitudinal drive}

Let us consider first the case of  longitudinal drives, $A_y=0$. In
Fig. \ref{F2}(a) we plotted $N(T)$ for different values of $\Omega$
at constant $D$. It is immediately apparent that the $N(T)$ peaks
are centered around
\begin{equation}
\label{TnA}
T_n=\left (n - \frac{1}{2}\right )\;T_\Omega, ~~~n=1,2,3, \dots,
\end{equation}
due to the fact that the left-to-right pore  crossings are most
likely to occur when the longitudinal ac force $A_x(t)$ points to
the right and vice versa. Such a synchronization mechanism has been
discussed at length in the SR literature \cite{RMP}. A detailed
numerical analysis (shown in Fig.~2(a) for $n=1$) proves that the
height of (or, more precisely, the area enclosed under) the $n$-th
distribution peak increases with $\Omega$ up to an optimal value
$\Omega=\Omega_n$, and then decreases for higher $\Omega$. Note that
$\Omega_n$ is determined by the optimal synchronization condition
\cite{gammaitoni}
\begin{equation}\label{bf}
\left (n - \frac{1}{2}\right)~T_\Omega=\tau_\Delta(D), ~~~n=1,2,3, \dots.
\end{equation}
Note that for $n=1$ this equality coincides with the spectral SR
condition in Eq.~(\ref{Dmax}). This proves that the first peak (but
this conclusion actually applies to all them!) can be enhanced to a
maximum by acting either upon $T_\Omega$ or $\tau_\Delta$. Hence,
geometric SR is also a {\it bona-fide} resonance \cite{gammaitoni}.

\subsection{Transverse drive}

We now consider the case of transverse drives, $A_x=0$. Clearly, a
drive parallel to the compartment wall of the cavity in
Fig.~\ref{F1}(a) cannot break the mirror symmetry of the system in
the $x$ direction. However, it impacts the diffusion process through
the pore, in that it presses the Brownian particle periodically
against the horizontal walls, twice a period. Pressed against the
top and the bottom walls of the cavity, symmetrically placed with
respect to the pore, the particle is less likely to escape the side
of the cavity it is trapped in; it is more likely to do so as
$A_y(t)$ reverses sign, twice a period. Therefore, one expects
$N(T)$ to develop a denser peak structure with
\begin{equation}
\label{TnB}
T_n=\frac{n}{2}\;T_\Omega, ~~~n=1,2,3, \dots,
\end{equation}
as shown in Fig.~\ref{F2}(b). Analogously to the situation of
Fig.~\ref{F2}(a), the distribution peaks become sharpest for an
optimal value of $\Omega$. An explicit numerical analysis (not shown) confirms that such an optimal synchronization occurs also under the resonance condition of Eq. (\ref{bf}), but with the difference that now $\tau_\Delta(D)$ stays for the transverse diffusion time $\tau_y=y_L^2/3D$. Although the right and left flows through the pore are modulated in
time, we stress that in the presence of a transverse drive $A_y(t)$,
no geometric SR can occur because of the persistent mirror symmetry
of the longitudinal motion.

\begin{figure}[bp]
\centering
\includegraphics*[width=8.0truecm]{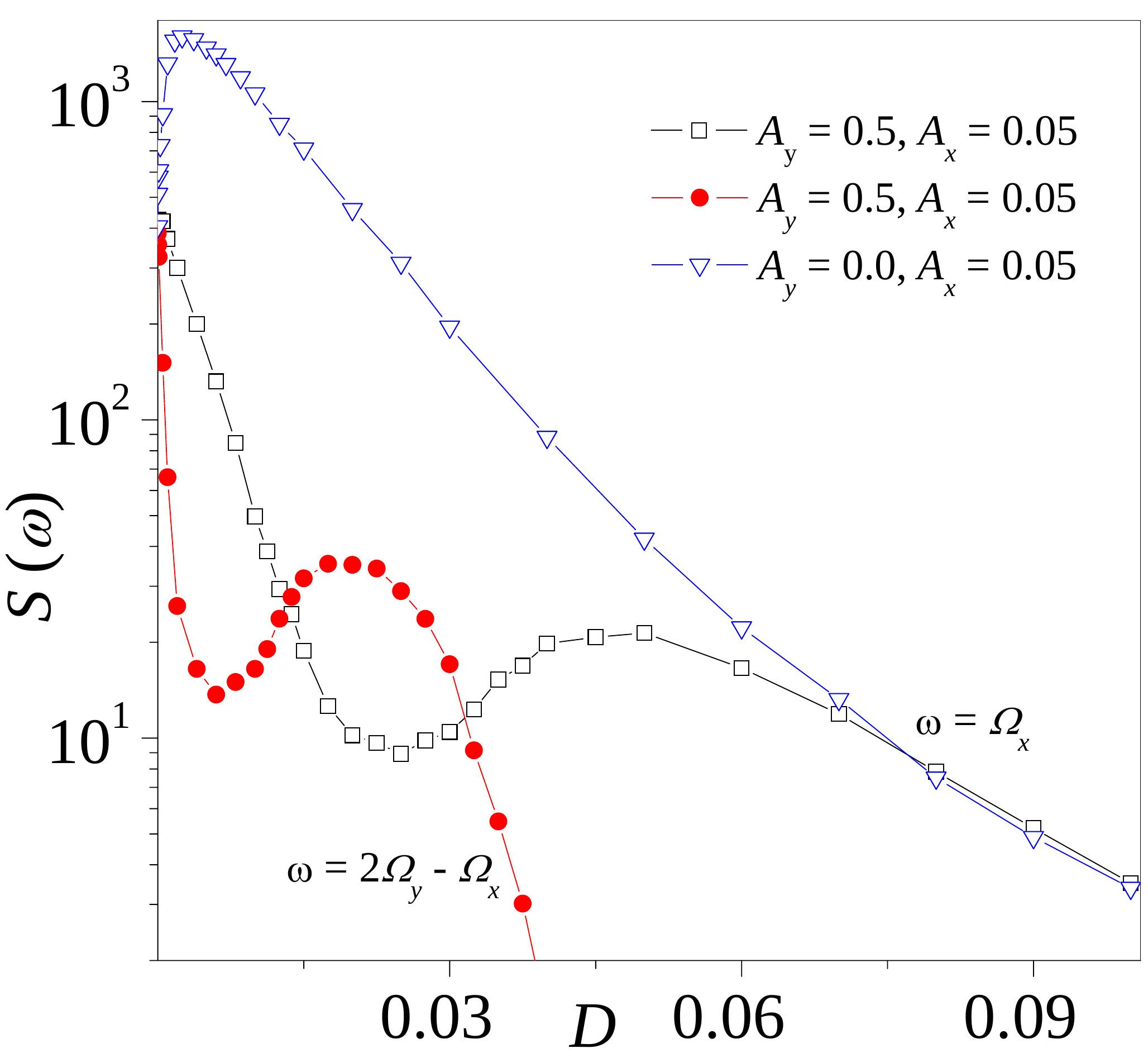}
\caption{(Color online) Resonant behavior of the two  lowest order
harmonic mixing components (1,0), (-1,1). For the sake of comparison
$S(\omega_{1,0})$, is shown also for $(A_x,A_y)=(0.005,0)$,
geometric SR. All simulation parameters here are as in
Fig.~\ref{F3}(b). \label{F4}}
\end{figure}

\section{Harmonic mixing} \label{HM}

The spectral characterization of geometric SR, as outlined in Sec.
\ref{gSR}, can provide  us with more insight into the resonant
transport mechanisms at work in a partitioned cavity. The analysis
of Ref. \cite{PRLgSR} is equivalent to taking the power spectral
density (PSD) of $x(t)$, $S(\omega)$, and evaluating the
$\delta$-like spike, $(\pi/2)\bar x^2(D) \delta(\omega-\Omega)$,
corresponding to the harmonic component of $\langle x(t)\rangle$
with frequency $\Omega$. As clearly shown in Fig.~\ref{F3}(a),
$S(\omega)$ develops a series of spectral spikes at
\begin{equation}
\label{omegan}
\omega_n=(2n-1)\,\Omega, ~~~n=1,2,3, \dots.
\end{equation}
As well known from the SR literature \cite{RMP}, the even harmonics
of the driving frequency are absent as a consequence of the $x \to
-x$ symmetry of the 2D Langevin equation (\ref{le}). On the other
hand, as anticipated in the final paragraph of Sec.~\ref{sync}, SR
cannot be induced by transverse drives. As expected, the PSD for
$A_x=0$ and $A_y \neq 0$ displayed in Fig. \ref{F3}(a) exhibits no
resonance spike.

\begin{figure}[bp]
\centering
\includegraphics*[width=8.5truecm]{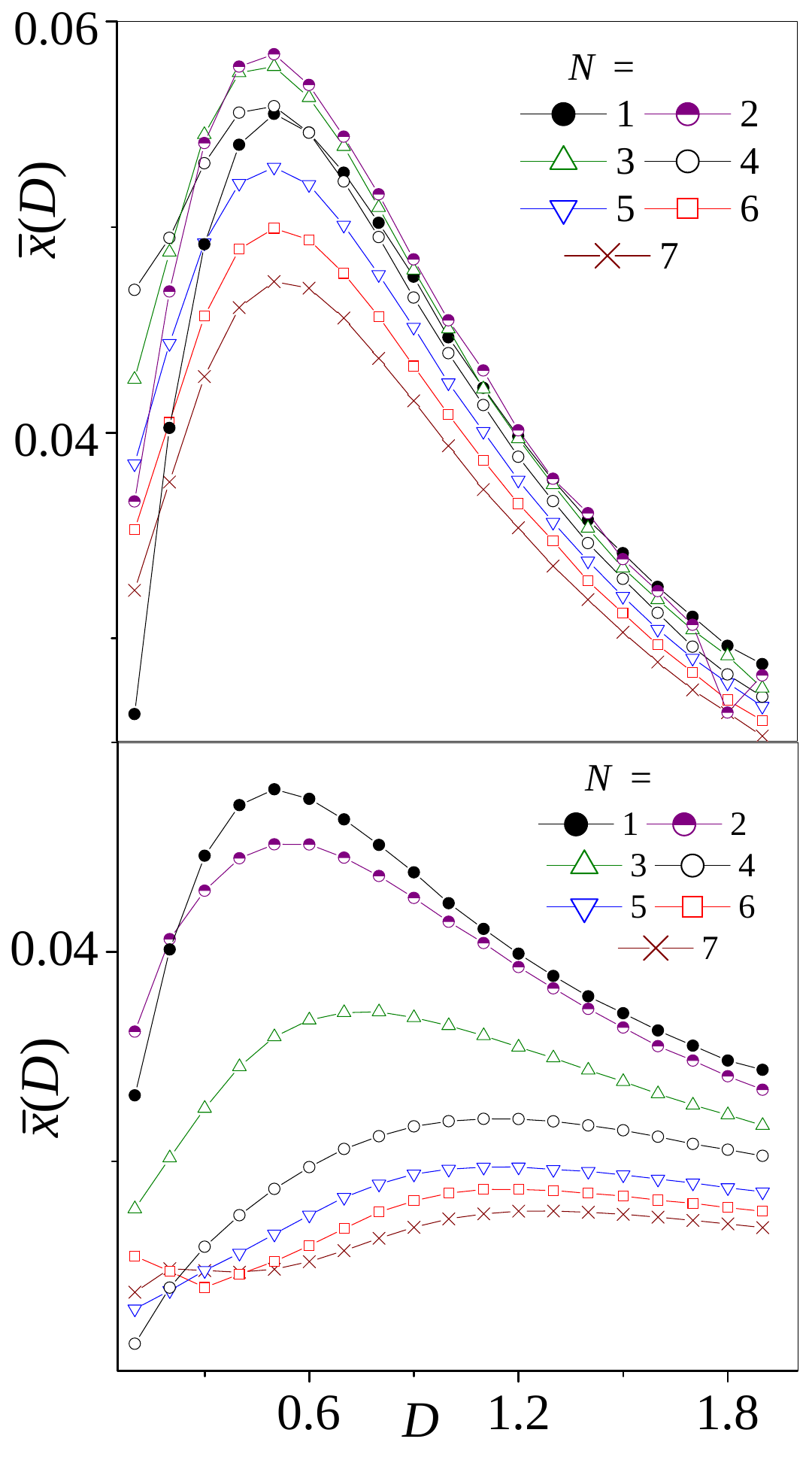}
\caption{(Color online) Geometric SR for interacting particles.
$\bar x(D)$ versus $D$ for the double cavity shown in Fig.~\ref{F1}
for $N$ identical particles repelling via the pair force of Eq.
(\ref{pot}) with $\alpha =1$ (a) and $6$ (b). The drive parameters
are $A_x=0.05$, $A_y=0$, and $\Omega_x=0.01$. \label{F5}}
\end{figure}

More interesting is the situation of Fig.~\ref{F3}(b), where the
Brownian particle is subjected to drives in both directions. The
intrinsic nonlinearity of the longitudinal and transverse flows
discussed in Sec. \ref{sync}A and B makes a mixing between them
possible, a phenomenon called  harmonic mixing (HM).

As discussed in Ref.~\cite{HM}, key ingredients for HM in a 1D
nonlinear system are: (i) nonlinearity of the driven process; (ii)
combination of at least two harmonic drives with angular frequencies
$\Omega_1$ and $\Omega_2$; (iii) commensuration of the driving
frequencies, that is, $\Omega_1/\Omega_2=p/q$, with $p$ ad $q$
relative prime numbers. Under such conditions, the system response
to the external drives develops a hierarchy of harmonics at
\begin{equation}
\label{omegamn}
\omega_{m,n}=m\Omega_1+n\Omega_2 ~~~m,n=0,\pm 1,\pm 2,\dots.
\end{equation}
Dynamical symmetries peculiar to the system can lead to the
suppressions of subsets of the harmonics $\omega_{m,n}$.

The system investigated in this section, however, is two
dimensional,  which means that HM can occur for any ratio of the
driving frequencies. In our simulations we employed orthogonal
harmonic drives, $A_x(t)$ and $A_y(t)$, with incommensurate
frequencies, $\Omega_x$ and $\Omega_y$. We know from Sec.~\ref{sync}
that the longitudinal flows driven by the transverse force alone can
only resonate at the {\it even} harmonics of $\Omega_y$, namely for
$\omega=2n\Omega_y$, $n=1,2,3, \dots$. This is an effect of the
mirror symmetry of the cavity with respect to the horizontal axis
passing through the center of the pore. Moreover, as mentioned
above, the mirror symmetry of the cavity with respect to its
compartment wall restricts the periodic components of $\langle x(t)
\rangle$ to the {\it odd} harmonics of $A_x(t)$, i.e.,
$\omega=(2n-1) \Omega_x$, $n=1,2,3 \dots$.

In conclusion, the HM spectrum of the longitudinal flow through the
cavity pore is expected to be
\begin{equation}
\label{omegaHM}
\omega_{m,n}=m\Omega_x+2n\Omega_y ~~~
\end{equation}
with $m=\pm 1, \pm 3, \pm 5, \dots$ (odd) and $n=0,\pm 1,\pm
2,\dots$,  with no commensuration condition on the ratio
$\Omega_x/\Omega_y$. Our prediction is corroborated by the PSD.
curves plotted in Fig.~\ref{F3}(b): $\Omega_x/\Omega_y$  is an
irrational number and, still, all detectable spectral peaks could be
identified by a pair of indices $(m,n)$ according to Eq.
(\ref{omegaHM}).

The practical implications of the HM of longitudinal and transverse
drives is that, while the time modulation introduced by the harmonic
signal $A_y(t)$, alone, cannot be picked up by the longitudinal
current across a pore, adding a small longitudinal signal makes
$A_y(t)$ detectable through the very same observable. We also stress
that mixing spikes with $n,m\neq 0$ are not necessarily small with
respect to the harmonics of the longitudinal signal, $n=0$. In fact,
all PSD spikes, $S(\omega_{m,n})$ manifest a SR dependence of their
own on the noise intensity. For instance, in Fig. \ref{F4}, for an
appropriate $D$ range, the HM harmonics $(-1,1)$ overshoots the
fundamental component $(1,0)$.

\begin{figure}[bp]
\centering
\includegraphics*[width=8.5truecm]{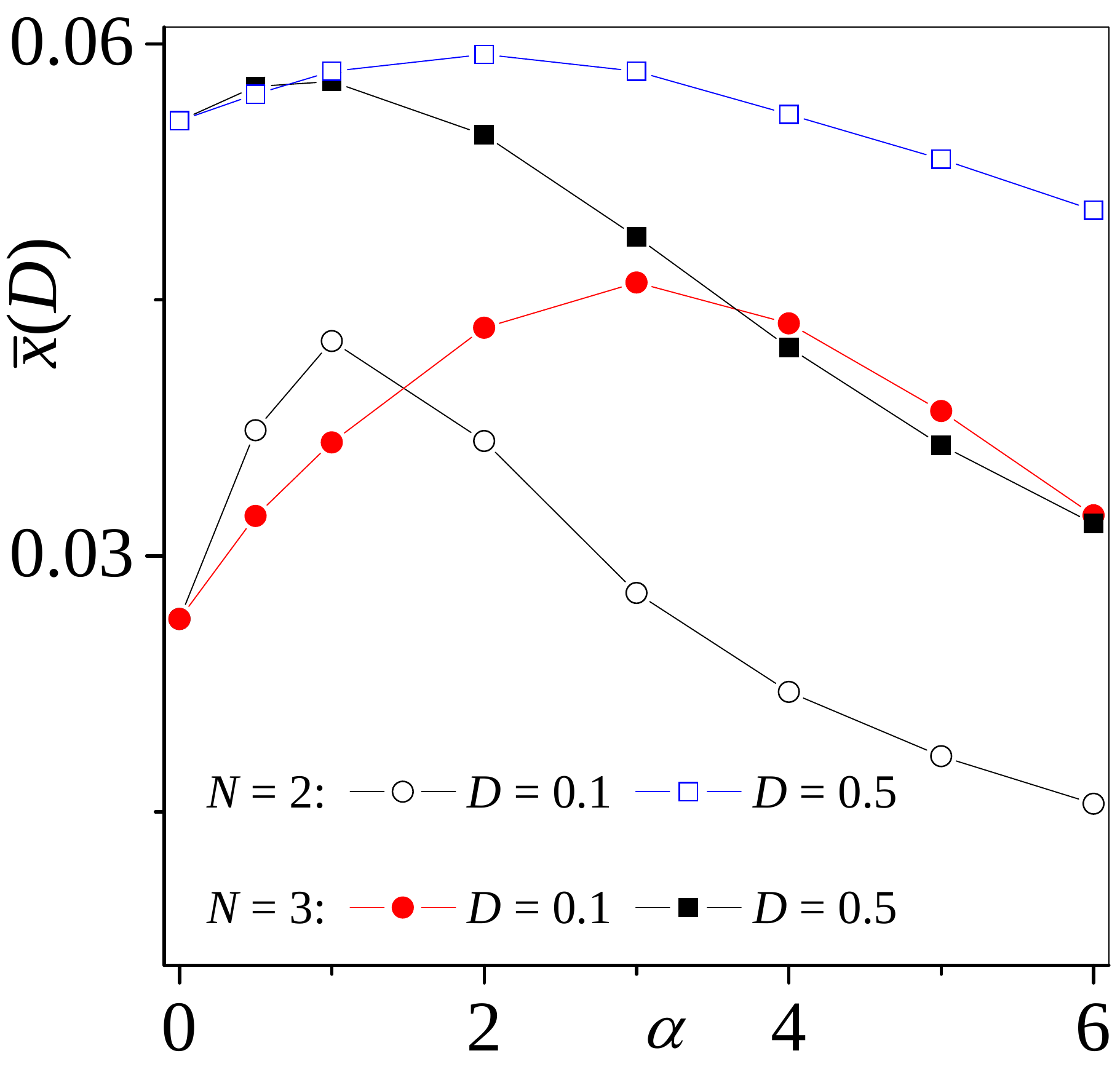}
\caption{(Color online) $\bar x(D)$ versus $\alpha$ for the same
setup as in Fig. \ref{F5} with $N$ and $D$ fixed (see legend). The
parameter $\alpha$ is the strength of the particle-particle
interaction. $\bar{x} (D)$ shows a peak for relatively small values
of $\alpha$, and slowly decreases for laser values of $\alpha$.
\label{F6}}
\end{figure}
\section{Particle interaction} \label{interaction}

To further explore the effects of the confining geometry on the
pore crossing mechanism, we consider now the role of particle
interaction. We assume that $N$ Brownian particles are randomly
distributed between the two cavities of the system and repel one
another through the standard vortex-vortex potential \cite{vortex},
\begin{equation}
\label{pot}
f_{i,j}=\frac{\alpha}{|\vec{r_i}-\vec{r_j}|},
\end{equation}
where $i,j=1,2, \dots, N$ and $i\neq j$. The situation modeled  here
is recurrent in biological physics
\cite{natural,Zwanzig,chemphyschem,PRLgSR,bere1,bere2}, where
constrained geometries often accommodate controllable concentrations
of suspended particles. The exact form of $f_{i,j}$ is not relevant
to the present discussion \cite{int_ratchet}. Here we address
the dependence of the geometric SR on the parameters $N$
(concentration) and $\alpha$ (coupling) of a system of interacting
particles.

Let us consider the geometry SR setup of Sec. \ref{gSR} with a
longitudinal harmonic drive, $A_x(t)$, of fixed angular frequency
$\Omega_x$. In Fig.~\ref{F5} we plot the curves $\bar x(D)$ for low
$\alpha$, in panel (a), and large $\alpha$, in panel (b), and
increasing particle concentrations. For low $\alpha$, increasing $N$
amounts to reducing the effective cavity volume accessible to the
Brownian particles. Moreover, when the particle coupling is
relatively weak, their interactions only exert a mean field effect
on the pore crossing process. As a consequence, one expects that, in
the regime of low couplings, the SR noise intensity, $D_{\rm max}$
in Eq.~(\ref{Dmax}), weakly depends on $N$, whereas the height of
the SR peak, Eq.~(\ref{xmax}), diminishes with raising $N$. This
description interprets qualitatively well the results of Fig.
\ref{F5}(a). An exception is represented by the curve with $N=1$,
which lies under the curve with $N=2$ and peaks at higher $D$. All
the remaining curves are centered around the same $D_{\rm max}$,
their maxima slowly decreasing for increasing $N$, with $N>2$. As a
matter of fact, when passing from $N=1$ to $N=2$, pair repulsion
clearly makes pore crossing more effective than for a single
particle. However, as argued above, this effect becomes negligible
when adding one particle at larger $N$.

For large $\alpha$, as shown in Fig.~\ref{F5}(b), the pair repulsion
can grow so strong that pair crossings become unlikely. Cavity
switching can only be achieved by pumping more noise into the
system, which means that the SR peaks must shift to higher $D_{\rm
max}$ for larger concentrations. Of course, for very large $N$, the
harmonic component of $\langle x (t) \rangle$ is suppressed
independent of the value of  $\alpha$.

The $\alpha$ dependence of the driven flow across the pore is
further  illustrated in Fig.~\ref{F6} for $N=2$ and $N=3$. The
curves for low noise, say, $D<D_{\rm max}$ for $\alpha=0$, are
particularly suggestive. The noise is so small that in the absence
of interactions pore crossing happens with a time constant
substantially larger than a half-drive period, see Eq. (\ref{Dmax}).
When switching on the interaction, the accessible free volume per
particle shrinks, which facilitates the escape of single particles
through the pore. This explains the raising branch of the curves
$\bar x$ versus $\alpha$. As the coupling $\alpha$ grows stronger,
particle repulsion gets global, with the consequence that particles
tend to move in deformable clusters (plastic flow \cite{nori}); at
exceedingly large $\alpha$, cavity switching stops and $\bar x$
decays to zero. According with the above argument, the resonant
dependence of $\bar x$ on $\alpha$ is less appreciable for higher
noise levels.

\section{Conclusions} \label{conclusion}

We have shown how particles suspended in a partitioned cavity can
diffuse across a pore or hole in the dividing wall, subjected to the
combined action of thermal fluctuations and period drives. The
particle flow across the pore is time-modulated at the drives'
frequencies, with amplitudes that can be optimized by controlling
the temperature of the system. This is a geometric effect where the
resonance condition depends on the shape of the cavity and on the
interactions among the particles it contains.

The mechanism discussed in this paper clearly does not depend on the
dimensionality of the cavity (experiments, e.g., on colloidal
systems, can more easily be carried out in 3D geometries), but can
be affected by other competing effects: (i) Pore structure; For a
finite-size particle, say, a translocating molecule, the actual
crossing time varies with the wall structure inside the pore and in
the vicinity of its opening \cite{natural}; (ii) Microfluidic
effects: The flow of the electrolytic suspension fluid across the
pore generates inhomogeneous velocity and electrical fields which
act on drift and the orientation of the driven particles
\cite{chemphyschem}. System-specific effects (i) and (ii) can in
principle be incorporated in our model by adding appropriate
potential terms to the Langevin equation (\ref{le}).

\acknowledgments We thank the RIKEN RICC for providing computing
resources. FN acknowledges partial support from the Laboratory  of
Physical Sciences, National Security Agency, Army Research office,
DARPA, Air Force office of Scientific Research, National Science
Foundation grant No.~0726909, JSPS-RFBR contract No. 09-02-92114,
Grant-in-Aid for Scientific Research (S), MEXT Kakenhi on Quantum
Cybernetics, and Funding Program for Innovative Research and
Development on Science and Technology (FIRST).  FM acknowledges
partial support from the Seventh Framework Programme under grant
agreement n° 256959, project NANOPOWER.

\end{document}